\documentclass[prb,aps,draft,tightenlines,twocolumn,
showpacs,floatfix]{revtex4}

\usepackage{graphicx}
\usepackage{dcolumn}
\usepackage{bm}
\usepackage{psfig}
\begin{document}
\small

\title{Spin-dependent transport in  lateral periodic magnetic
 modulations: a scheme for spin filters}
\author{J. Zhou}
\affiliation{Structure Research Laboratory, University of Science and
Technology of China, Academia Sinica,  Hefei, Anhui, 230026, China\\
Department of Physics, University of Science and
Technology of China, Hefei, Anhui, 230026, China}
\altaffiliation{Mailing Address.}
\author{Q. W. Shi}
\affiliation{Structure Research Laboratory, University of Science and
Technology of China, Academia Sinica,  Hefei, Anhui, 230026, China\\
Department of Physics, University of Science and
Technology of China, Hefei, Anhui, 230026, China}
\author{M. W. Wu}
\thanks{Author to whom correspondence should be addressed}%
\email{mwwu@ustc.edu.cn}%
\affiliation{Structure Research Laboratory, University of Science and
Technology of China, Academia Sinica,  Hefei, Anhui, 230026, China\\
Department of Physics, University of Science and
Technology of China, Hefei, Anhui, 230026, China}

\date{\today}

\begin{abstract}
A scheme for spin filters is proposed by studying the coherent transport
of electrons through quantum wires with lateral magnetic modulations.
Unlike other schemes in the literature, the modulation in our scheme is much weaker than the Fermi energy. 
Large spin polarization through the filter is predicted.
Further study suggests the robustness of this spin filter.
\end{abstract}
\pacs{85.75.-d, 73.23.Ad, 72.25.-b}
\maketitle

The realization of spintronic devices relies on the ability to inject
a spin-polarized current into a semiconductor.\cite{prinz} 
Progress has been made in injecting polarized electrons from 
ferromagnets\cite{ohno,zhu,hanb} or semimagnetic 
semiconductors\cite{fied,mala,ghali} into
semiconductors. Besides these efforts, generating spin polarization (SP)
through a spin filter has aroused more and more attention.\cite{filter} 
In these works, spin-selective {\em barriers} or stubs\cite{stub} 
are essential to realize the SP.
Other method such as generating SP
inside semiconductors by reflection at the interface with a
ferromagnet has also been proposed.\cite{lu} 
In this letter, we propose a scheme for a spin filter where the SP
is generated during the transport without tunneling
through any barrier or being mode-selected by any stub. 

We consider the electron ballistic transport through a semiconductor nanowire
under the periodic spin dependent modulation shown in Fig.\ 1.
Here the spin dependent potential
has the Zeeman-like form: $V_\sigma(x)=\sigma V_0g(x)$ 
with $g(x)=1$ if $x$ is located at the $A$-layer, and 
0 otherwise.  $\sigma$ is $\pm1$ for spin-up and -down electrons 
respectively. $V_0$ denotes a spin-independent parameter for 
the strength of the potential. 
Therefore, spin-up and -down electrons experience different periodic 
potentials: spin-up electrons coherently transport under the modulation 
of periodic barriers
while spin-down ones under the modulation of periodic wells. 
The transmission and reflection coefficients can be easily obtained 
and are spin dependent. Some earlier schemes for spin filters are also 
reported based on similar spin dependent
modulation.\cite{filter}  However, they only work in the 
assumption of the large potential  $V_0>E_F$ with
$E_F$ representing the Fermi energy. 
This is because a large barrier for the spin-up electrons
strongly suppresses the transmission of spin-up electrons,  therefore
the spin-up current decays exponentially. Nevertheless the spin-down electrons 
can easily transmit through the large wells. Generally speaking, the spin 
dependent modulation is weak except in the extreme conditions such
as applying a strong magnetic field or using a heavily-doped ferromagnetic
semiconductor which still remains a challenge and
will  cause new problems that limit the application. In our scheme, we
focus on the weak modulation case, {\em i.e.},  $V_0/E_F\ll 1$. 
In this case, even spin-up electrons do not see any true barriers but 
rather ``transparent'' barriers (TB's). Therefore 
a single TB (well) affects the transmission coefficient of 
a spin-up (-down) electron only a little bit. Hence, the SP is
negligible when the spin-up/down electrons coherently transport 
through a single TB/well. However, when electrons transmit
through a set of weakly periodic TB's or wells, under right conditions,
a new feature appears: electrons with different spin may pick up
these small SP's and accumulate to a large one
after they transmit through a large number $N$ of TB's (wells).
Moreover, we find the spin polarization shows oscillations with $N$.
Furthermore, 100\ \% SP is also predicted in our scheme. 
This is because with the periodic modulations, there exists
an energy gap. The positions of the gap for the spin-up and -down
electrons are separated because the modulations are different.
Hence, when the Fermi energy of the leads is within the gap regime 
of the spin-up (-down) electrons, the 
transmission coefficient for spin-up (-down) will decay 
exponentially with $N$  while that for the spin-down (-up) 
still  oscillates  with $N$.

\begin{figure}[h]
\vskip-0.1cm
\centering
  \psfig{figure=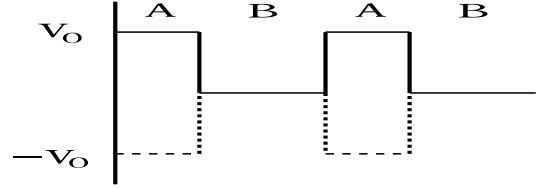,width=7.cm,height=2.5cm,angle=0}
\vskip-0.3cm
  \caption{The modulation for spin-up (solid curve) and spin-down
    (dashed curves) electrons.}
\end{figure}

We describe a quantum wire along the $x$-direction with the 
tight-binding approach.\cite{Mireles}
By taking a two dimensional grid with $N_y$ grid sites along the
transverse direction and $N_x$ sites along the wire, we have
\begin{eqnarray}
H&=&\sum_{l,m,\sigma} \epsilon_{l,\sigma} c^\dagger_{l,m,\sigma}
c_{l,m,\sigma} + 
t \sum_{l,m,\sigma} (c^\dagger_{l+1,m,\sigma}c_{l,m,\sigma} \nonumber \\
&&+ c^\dagger_{l,m+1,\sigma}c_{l,m,\sigma} +h.c.)\ ,
\end{eqnarray}
where  $l$ and $m$ denote the  coordinates along the  $x$- and $y$-axis
respectively.
$\epsilon_{l,\sigma}=\epsilon_0+\sigma V_0$ ($=\epsilon_0$)
when $l$ locates at the A (B) layer, denotes the on-site energy
with $\epsilon_0=-4t$.
 $t=-\hbar^2/2m^\ast a^2$ is the hopping energy with $m^\ast$ 
and $a$ standing for the effective mass and  the ``lattice'' constance
respectively. 

The spin dependent conductance is calculated using  the
Landauer-B\"{u}ttiker\cite{Bu} formula with the help of 
the Green function method.\cite{Da} The two-terminal spin-resolved 
conductance is given by
$G^{\sigma \sigma^\prime}=(e^2/h)\mbox{Tr}[\Gamma^{\sigma}_{1}
G^{\sigma\sigma^\prime+}_{1N_x}\Gamma^{\sigma^\prime}_{N_x}G^{\sigma
^\prime\sigma -}
_{N_x1}]$ with  $\Gamma_{1(N_x)}$ representing the
self-energy  function for the isolated ideal leads.\cite{Da} 
We choose the  perfect ideal ohmic contact between the
leads and the semiconductor. $G^{\sigma\sigma^\prime+}_{1N_x}$ and 
$G^{\sigma\sigma^\prime-}_{N_x1}$ are the 
retarded and advanced Green functions
for the conductor, but with the effect of the leads included. 
The trace is performed over the spatial degrees of freedom along the 
$y$-axis. Without the spin-flip process, 
one can define the SP as 
$P=(G^{\uparrow \uparrow }-G^{\downarrow \downarrow })/
(G^{\uparrow \uparrow }+G^{\downarrow \downarrow })$.

\begin{figure}[htb]
\vskip-0.3cm
  \psfig{figure=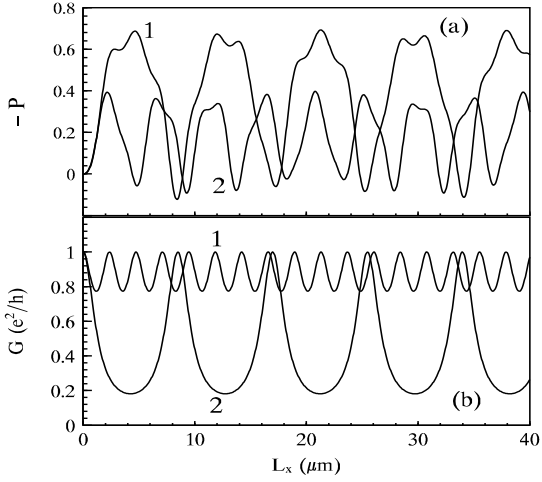,width=9.5cm,height=7.cm,angle=0}
\vskip-0.3cm
  \caption{(a) Spin polarization {\em vs.} the length of the filter.
Curve 1 and 2 correspond to two different modulations for Case I and
II. (b) Spin dependent conductance $G^{\downarrow\downarrow}$ (curve 1) 
and $G^{\uparrow\uparrow}$ (curve 2) {\em vs.} the filter length
for Case I.}
\end{figure}

We perform a numerical calculation for a quantum wire with
fixed width $N_y=40a$. The  hard wall potential is applied in 
this transverse direction which makes the lowest energy
of the $n$-th subband (channel) to be $\varepsilon_n=
2|t|-2|t|\cos[n\pi/(N_y+1)]$.
 $a=20$~{\AA} 
throughout the computation.
The total length of a single unit (an A-layer plus a B-layer) 
is fixed at $30a$. We take the Fermi 
energy $E_f=0.01697|t|$ and the Zeeman splitting energy
$V_0=0.001|t|$. Such a choice of the Fermi energy guarantees 
not only the lowest subband (single mode)  contribute to the conductance
but also gives the SP. 
It is noted that $V_0/E_f \sim 0.06$ is very small.

In Fig.\ 2(a) the SP $P$ is plotted as a function
of the length of the semiconductor wire $L_x$ for two
different modulations:  Case I, the length of the A-layer $L_A$ is the same
as that of the B-layer $L_B$, (curve 1); Case II,
$L_A=17a$ (curve 2). $L_A+L_B$ is always fixed to be $30a$.
It is seen from the figure that unpolarized injection from the left
lead acquires SP when it reaches to the right one
if the length of the modulation is long enough. When $L_x$ is around
1\ $\mu$m, the SP's for the two cases all reach to 10~\%.
Oscillations appear when the filter length further increases.
The maximum SP differs for different modulations. 
For case I, nearly 70\ \% SP can be achieved when $L_x$ is
around 4\ $\mu$m.
In order to understand this SP oscillation,
we plot the spin dependent conductances $G^{\uparrow \uparrow}$ 
and $G^{\downarrow \downarrow}$
versus the filter length $L_x$ in Fig.\ 2(b) for Case I. 
It is clearly observed that both conductances oscillate with $L_x$,
nevertheless with different periods. 
The period for spin-up conductance is around 8.6\ $\mu$m and
that for spin-down one is about 2.4\ $\mu$m. Therefore, through several 
periods, large mismatch accumulates and the peak SP is 
reached when the position of the peak 
transmission of one spin  is around the position of the 
valley transmission of the opposite spin.

\begin{figure}[htb]
\psfig{figure=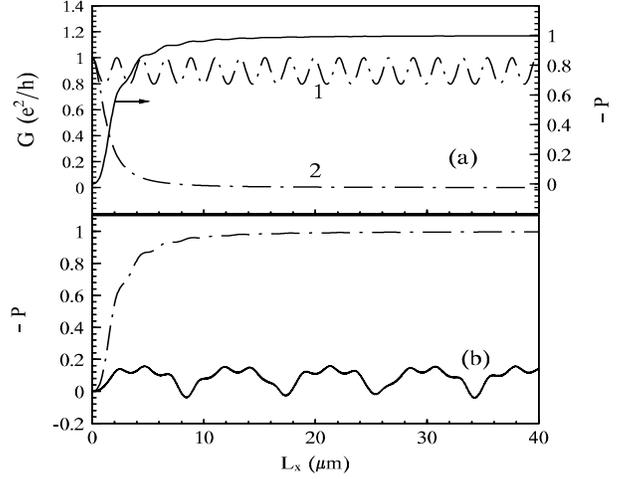,width=9.5cm,height=7.cm,angle=0}
\vskip-0.3cm
\caption{(a) Spin dependent conductance $G^{\downarrow\downarrow}$ (curve 1) 
and $G^{\uparrow\uparrow}$ (curve 2) as well as 
SP {\em vs.} the length of the filter
for the Case I, but with the change of the Fermi energy. 
It is noted that the scale of 
the SP is at the right side of the figure.
(b) SP {\em vs.} the filter length for the same modulation as in (a).
Chain curve: single mode; Solid curve: three modes.}
\end{figure}

To further elucidate the effect of the SP oscillation,
we  consider an exact one dimensional scattering problem
of an electron with large kinetic energy $E_F$ passing through
the same spin dependent modulations as shown in Fig.\ 1. 
After $N$-units, 
the transmission coefficient is 
\begin{equation}
T_\sigma(N)=\{1 + [|\gamma_\sigma|^2 - \sin^2(\delta\theta
_\sigma)]
\sin^2(N\delta\theta_\sigma)/\sin^2(\delta\theta_\sigma)\}^{-1}
\end{equation}
 with 
$\delta\theta_\sigma={\arccos[-\cos(\kappa_\sigma 
L_A+ k_FL_B)+1/2 \sin(\kappa_\sigma L_A)}\linebreak
{\sin(k_FL_B)(\kappa_\sigma/k_F
+k_F/\kappa_\sigma-2)]}$. 
$k_F=\sqrt{2m^\ast E_F/{\hbar^2}}$ 
and
$\kappa_\sigma=\sqrt{2m^\ast (E_F+\sigma V_0)/\hbar^2}$
are the free electron momentum at B-layers  and the electron momentum
under the potential of the TB's or wells at A layers  respectively.
$\gamma_\sigma =i[\sin(k_FL_B)\cos(\kappa_\sigma
L_A)+\sin(\kappa_\sigma L_A)\cos(k_FL_B)(\kappa_\sigma/k_F
+k_F/\kappa_\sigma)/2]$
is a pure imaginary  number.
Equation (2) clearly shows that $T_\sigma(N)$ is a  periodic 
function of $N$. When $N\delta\theta_\sigma = m_1\pi $ or 
$N\delta\theta_\sigma=(m_2+1/2)\pi$ is satisfied 
($m_1$ and $m_2$ here represent
integers), the peak or valley appears respectively with 
the value of the peak and the valley being 1 and $\sin^2
(\delta\theta_\sigma)/|\gamma_\sigma|^2$.
The length corresponding to the first large SP is determined
approximately by matching the peak of one spin with the valley
of the opposite one:
$[|m_1-m_2+1/2|\pi/(|\delta\theta_\uparrow - \delta\theta_\downarrow|)]
(L_A+L_B)$ by choosing the smallest $m_1$ and $m_2$ to satisfy
$m_1/ \delta\theta_\uparrow \sim (m_2+1/2)/ \delta\theta_\downarrow$.
In order to have large SP, the oscillation of each
spin transmission should be large  enough. The oscillation amplitude of 
each spin transmission can be determined by subtracting the valley
transmission from the peak one:
$1-\sin^2(\delta\theta_\sigma)/|\gamma_\sigma|^2
=[(\kappa_\sigma/k_F-k_F/\kappa_\sigma)^2\sin^2(\kappa_\sigma
L_A)/4]/[\sin(\kappa_\sigma L_A+k_FL_B)+\sin(\kappa_\sigma
L_A)\cos(k_F L_B)(\kappa_\sigma/k_F+k_F/\kappa_\sigma-2)/2]^2$.
As $V_0/E_F$ is very small and therefore
$\kappa_\sigma/k_F+k_F/\kappa_\sigma-2$  is a small number, it is
clear that only if one chooses the Fermi
energy $E_F$ to  satisfy the  
condition $k_F(L_A+L_B) \sim \pi$, one can get large 
transmission oscillation and hence large SP.
Facilitated with this simplified model, one may understand the
features in Fig.\ 2: When choosing the parameters of the case I, 
$\delta\theta_\uparrow$ and $\delta\theta_\downarrow$ are around 0.022
and 0.08 respectively. Therefore the corresponding period of length
of spin-up (-down) is 8.6\ $\mu$m  (2.4\ $\mu$m) and
the first largest SP is estimated to be at 4.8\ $\mu$m. These numbers 
are the same as those in Fig.\ 2. The same is true for case II.

\begin{figure}[htb]
  \psfig{figure=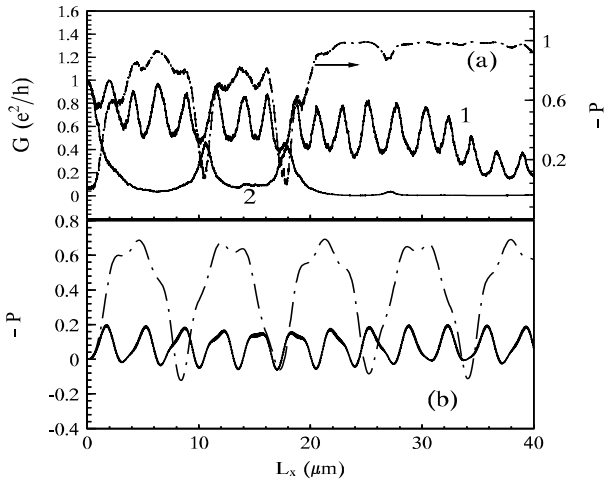,width=9.5cm,height=7.cm,angle=0}
\vskip-0.3cm
  \caption{(a) Spin dependent conductance $G^{\downarrow\downarrow}$ (curve 1) 
and $G^{\uparrow\uparrow}$ (curve 2) as well as 
SP {\em vs.} the length of the filter
for case I in the presence of a Anderson disorder ($W=0.01|t|$).
 It is noted that the scale of 
the SP is at the right side of the figure.
(b) SP {\em vs.} the filter length for the same modulation as in (a)
but without disorder. Chain curve: without Rashba effect; Solid curve: 
with Rashba effect.}
\end{figure} 

It is interesting to see that 100\ \%  SP can  be obtained if one chooses
$E_F=0.01701|t|$  as shown in Fig.\ 3(a).  When $L_x$ is longer than
10\ $\mu$m,  $P$ stays 100\ \%.  
This can be understood from Eq.\ (2): The
phase shift $\delta \theta _\sigma$ for 
spin-up electron is 0 while the one for spin-down electron is 0.085.
Therefore $T_\downarrow(N)=1/[|\gamma_\downarrow|^2N^2]$ which decays
to 0 when $N$ increases. The length period of the oscillations of 
transmission coefficient for spin-down electron is 2.2\ $\mu$m, exactly as 
shown in the figure.
If one further increases $E_F$ slightly, $\delta\theta_\uparrow$ 
becomes a pure imaginary number which implies that $E_F$ is 
within the regime of the gap of the spin-up electrons.  
As $\sin(i|\delta \theta_\uparrow|N)=i\sinh(|\delta
\theta_\uparrow|N)$,  from  Eq.(2) one can see that 
that $T(N)$ decays   exponentially  with $N$.
The gap can be determined by keeping  $\delta\theta_\sigma$ 
imaginary:  $[-\cos(\kappa_\sigma 
L_A+ k_FL_B)+1/2 \sin(\kappa_\sigma L_A)\sin(k_FL_B)
(\kappa_\sigma/k_F+k_F/\kappa_\sigma-2)]\geq 1$. 
To the second order of  $(V_0/E_F)$, the gap for the spin $\sigma$
is given by 
 [$\hbar^2[K_\sigma -{1\over 2\pi}{V_0 \over E^0_F}k^0_F 
\sqrt {\sin{(k^0_FL_A)}
\sin (k^0_FL_B)}]^2/(2m^\ast), 
\hbar^2[K_\sigma +{1\over 2\pi}{V_0 \over E^0_F}k^0_F 
\sqrt {\sin {(k^0_FL_A)}\sin (k^0_FL_B)}]^2/(2m^\ast)$] with 
$k^0_F=\pi/ \linebreak{(L_A+L_B)}$, $E^0_F= \hbar^2 (k^0_F)^2 /{2m^\ast}$ and 
$K_\sigma =  k^0_F(1+{\sigma\over 2}{V_0\over
  E^0_F}{L_B\over(L_A+L_B)})$ respectively.
Therefore, the gap of spin-up electrons
 is higher than that of the spin-down ones. It is noted that if
 $L_A=L_B$, the width of the gap is largest.
 
We also investigate the multi-mode effect on the spin filter efficiency.  
In Fig.\ 3(b), the SP  is plotted as a function of the length $L_x$ when
$E_F=0.0637|t|$ which corresponds to the three-mode transport but 
with the condition $k_F(L_A+L_B)\sim\pi$ only satisfied for the third mode. 
The other parameters  are all the same as those
in Fig.\ 3(a). It is seen from the figure that the SP is still kept, 
but much smaller (still as high as 18\ \%). 
This is because the contribution to the SP only comes from the third mode. 

In order to check the robustness the spin filter proposed above, we now
include the Anderson disorder and  the Rashba effect in the single mode 
case to investigate their effects on the SP. In Fig.\ 4(a), 
the spin dependent  conductances $G^{\uparrow \uparrow}$ and 
$G^{\downarrow \downarrow}$ are plotted
against the filter length $L_x$ for case I, but with the 
Anderson disorder included. 
The strength of the disorder is taken to be $W=0.01|t|$, 
one order of magnitude larger than $V_0$.
It is found that the disorder makes the transmission
coefficients decay with $L_x$. However the
SP is {\em increased} to $\sim 100$\ \% when the filter length is 
longer than 20\ $\mu$m. 
This is because  spin-up electrons decay much faster when
$E_F$ is close to the  bottom of the gap.
We also calculate the SP in the presence of the 
Rashba effect\cite{parrek} $\lambda \sum_{l,m,\sigma
  \sigma^\prime}[c^{\dagger}_{l+1,m \sigma }c_{l,m,\sigma^\prime}
(i\sigma_y^{\sigma\sigma^\prime})-c^\dagger_{l,m+1,\sigma }
c_{l,m,\sigma }(i\sigma_x^{\sigma\sigma^\prime})]$ with
$\lambda=0.002|t|$  in Fig.\ 4(b).\cite{note} It is seen 
that the SP is reduced by the Rashba 
term as the later mixes the spins, but still with a noticeable polarization. 

In summary, we have proposed a  scheme for spin filters by studying 
the coherent transport of electrons through quantum wires with the 
lateral magnetic modulations. Unlike other schemes in the literature,
the modulation in our scheme is much {\em weaker} than the Fermi
energy.  A large SP is predicted if the  condition $k_F(L_A+L_B) \sim \pi$
is satisfied. Further study also shows 
the robustness of this scheme. 
The magnetic modulation can be realized by sticking the
magnetic strips on top of the sample or using magnetic 
semiconductor as A-layer.

MWW is supported by the  ``100 Person Project'' of Chinese Academy of
Sciences and Natural Science Foundation of China under Grant No.
90303012 and 10247002. He would like to 
thank valuable discussions with Bruce McCombe and  Hong Luo.


\begin{thebibliography}{10}
\bibitem{note} Here the SP is defined as 
$P=(G^{\uparrow\uparrow}+G^{\downarrow\uparrow}-
G^{\downarrow\downarrow}-G^{\uparrow\downarrow})/
(G^{\uparrow\uparrow}+G^{\downarrow\uparrow}+G^{\downarrow\downarrow}
+G^{\uparrow\downarrow})$.
\end{thebibliography}
\end{document}